\documentclass[prd,showpacs,showkeys]{revtex4}
\usepackage{amsmath,amsthm,amsfonts,amssymb}
\usepackage[mathcal]{eucal}
\usepackage{mathrsfs}
\usepackage{graphicx}
\usepackage{dcolumn}
\usepackage{latexsym}
\usepackage{epsfig}
\usepackage{bm}
\usepackage[all]{xy}
\newcommand{\ie}{\begin{equation}}
\newcommand{\fe}{\end{equation}}

\def\text#1{\mbox{#1}}

\begin{document}

\title{{Tensor gauge field localization in branes}}
\author{M. O. Tahim,$^{a,b}$ W. T. Cruz,$^{a,c}$ and C. A. S. Almeida$^a$}
\affiliation{$^a$Departamento de F\'{i}sica - Universidade Federal do Cear\'{a} \\ C.P. 6030, 60455-760
Fortaleza-Cear\'{a}-Brazil}
\affiliation{$^b$Departamento de Ci\^{e}ncias da Natureza, Faculdade de Ci\^{e}ncias, Educa\c{c}\~{a}o e Letras
do Sert\~{a}o Central (FECLESC), Universidade Estadual do Cear\'{a}, 63900-000 Quixad\'{a}-Cear\'{a}-Brazil}
\affiliation{$^c$Centro Federal de Educa\c{c}\~{a}o Tecnol\'{o}gica do Cear\'{a} (CEFETCE), Unidade
Descentralizada de Juazeiro do Norte, 63040-000 Juazeiro do Norte-Cear\'{a}-Brazil}

\begin{abstract}In this work we study localization of a Kalb-Ramond tensorial gauge field on
a membrane described by real scalar fields. The membrane is embedded in an AdS-type five dimensional bulk space,
which mimic a Randall-Sundrum scenario. First, we consider a membrane described by only a single real scalar
field. In that scenarios we find that there is no localized tensorial zero mode. When we take into account
branes described by two real scalar fields with internal structures, we obtain again a non-localized zero mode
for a Kalb-Ramond tensorial gauge field. After modifying our model of one single scalar field by coupling the
dilaton to the Kalb-Ramond field, we find that this result is changed. Furthermore, we analyze Kaluza-Klein
massive modes and resonance structures.
\end{abstract}



 \maketitle

\section{Introduction}

In scenarios containing extra dimensions in the background of membranes a very important subject is related to
mechanisms of localization of several fields with different spins. Such importance is due to a very simple
motivation: based on this idea all of the characteristics of a lower dimensional effective model must be
obtained. Characteristics such that coupling constants and the masses of the fundamental particles depends on
the size of the extra dimension and this fact may give new insights in order to understand some problems
presented by the Standard Model.

There are at least two different scenarios of extra dimensions. The first one consider small and compact
dimensions whose effects in lower energies is not observable. The second one regards the extra dimensions as
having infinite size and they are not observable simply because we have not permission (an energy permission) to
"walk" along them. The idea is that we live inside a membrane embedded in a spacetime of dimensionality bigger
than $4$. It is interesting to compare the differences between these two scenarios in models of field
localization. For example, related to gauge vector fields of the Standard Model, it is well known that in a
scenario where the extra dimension is compact the localization of this kind of field is not favored due to
phenomenological constraints imposed by the Standard Model \cite{rizzo}. In the case where the extra dimension
is infinite there is not localization unless the gauge field couples to the dilaton field \cite{kehagias}. We
should try to understand the behavior of other sort of gauge fields in the context of membranes, in particular
those with higher spins. The question we ask here is: can these fields have observable effects? Regarding this
question, there is a new interest in theories containing gauge fields with higher spins. The basic reason for
that is its existence on anti-de Sitter spaces \cite{ads-gauge}. Such characteristic signals for their relevance
for AdS/CFT correspondence \cite{germani-kehagias}. String theory also gives additional support to higher spin
fields. Indeed, string theory contains an infinite number of higher spin fields with consistent interactions. In
the low-tension limit their masses disappear. Massless higher spin theories are thus the natural candidates for
the description for the low-tension limit of string theory at the semi-classical level. The hope is that the
understanding of the dynamics of higher spin fields could help towards a deeper insight of string theory, which
now is mainly based on its low-spin excitations and their low-energy interactions.

The main subject of this work is related to the Kalb-Ramond gauge field, a rank two antisymmetric field, which
is the simplest case in the list of higher rank fields to be studied. This field appears in effective theories
of specific low energy superstring models, as cited above, and can describe axion physics or torsion of a
Riemannian manifold. When interpreted as torsion, it is known that this field, when studied in the
Randall-Sundrum background, possesses a localized zero mode extremely suppressed by the size of the extra
dimension \cite{sengupta}. Such a result leads the authors of this reference to speculate about a spacetime
torsion in our universe, even a small one. On the other hand, in string theories the appearance of axions from
the antisymmetric tensor fields is quite natural \cite{witten}.

Specifically in this work we make an analysis of mechanisms of localization for the antisymmetric tensor gauge
field in a scenario where the extra dimension has infinite size. The method we follow is quite different from
that described in \cite{sengupta} since we study smooth AdS-like backgrounds. In order to impose such a
condition it is necessary to implement the membranes of the model in a more realistic way. For such, we use kink
defects embedded in a higher dimensional spacetime. In this way, we can avoid problems related to spacetime
singularities presented in the Randall-Sundrum idea.

We also make use of models containing several scalar fields whose solutions describe thin membranes and
membranes with internal structures \cite{dionisio-adauto}, including a scenario of the coupling of Kalb-Ramond
and dilaton fields. In particular, in this last scenario, we present results about zero and massive modes of the
Kalb-Ramond tensor field.

The organization of this work is as follows: in the second section solutions describing the gravitational
background due to smooth membranes are studied; in the following section the Kalb-Ramond field is included in
this scenario; in the fourth and fifth sections we introduce the dynamics of models containing two fields,
including the dilaton; in the sixth section, the coupling between the dilaton and the Kalb-Ramond field is
studied and finally, the last section is reserved to discussions about Kaluza-Klein massive modes.

\section{The kink as a membrane}

In this section we study the solutions for the Einstein's equations in the background of a single thin membrane.
The membrane is a kink embedded in a $D=4+1$ spacetime. We study a spacetime background solution preserving four
dimensional Lorentz symmetry. These solutions in general describe AdS-like spacetimes. The action for the model
is \cite{kehagias}:
\begin{equation}\label{h}
S=\int d^{5}x \sqrt{-G}[2M^{3}R-\frac{1}{2}(\partial\phi)^{2}-V(\phi)].
\end{equation}
In the action above the field $\phi$ generates the membrane of the model, $M$ is the Planck constant in $D=5$
and $R$ is the curvature tensor. The equation of motion for the field $\phi$  supports a kink solution even in a
gravitational background. For this case, the ansatz for the spacetime metric is:
\begin{equation}
ds^{2}=e^{2A(y)}\eta_{\mu\nu}dx^{\mu}dx^{\nu}+dy^{2}.
\end{equation}
Despite the warp factor $A(y)$, the metric above preserves $D=4$ Lorentz invariance. As usually, in this work a
capital index like M stands for 0, 1, 2, 3, 4 while Greek letters stands for 0, 1, 2, 3. In order to make the
model simpler we assume the field $\phi$ and the function $A$ as only dependent of the extra dimension $y$. The
resultant equations of motion coming from the action (\ref{h}) are \cite{kehagias}
\begin{eqnarray}
R_{MN}-\frac{1}{2}G_{MN}R=\frac{1}{4M^3}\{{\partial_M \phi \partial_N \phi- G_{MN}[\frac{1}{2}
(\partial\phi)^2+V(\phi)]}\}, \\\nonumber
\frac{1}{\sqrt{-G}}\partial_M\{\sqrt{-G}G^{MN}\partial_N\phi\}=\frac{\partial V}{\partial \phi}
\end{eqnarray}

\begin{figure}
\includegraphics[width=8cm]{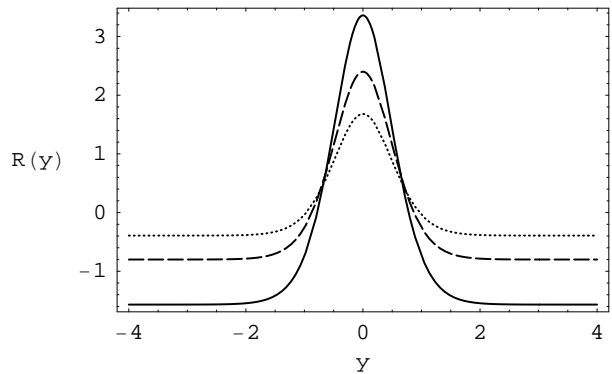}
\caption{
  \label{curv}   Plots of the curvature invariant $R(y)$ with
$\beta=0.07$ (dotted line), $0.1$ (dashed line), $0.14$ (solid line) and $a=1$.}
\end{figure}

For a situation without a gravitational background it is easy to show that, for the potential function
$V(\phi)=\frac{\lambda}{4}(\phi^{2}-v^{2})^{2}$, the kink modeling the membrane is given by $\phi(y)=v\tanh(ay)$
where $a^2=\frac{\lambda v^{2}}{2}$. This is a solution for the equation of motion for the scalar field. The
equations of motion for the case of the AdS-like gravitational background described above are given by:
\begin{equation}\label{mov1}
\frac{1}{2}(\phi^{\prime})^{2}-V(\phi)=24M^{3}(A^{\prime})^{2},
\end{equation}
and
\begin{equation}\label{mov2}
\frac{1}{2}(\phi^{\prime})^{2}+V(\phi)=-12M^{3}A^{\prime\prime}-24M^{3}(A^{%
\prime})^{2}.
\end{equation}
Note that the prime means derivative in respect to the extra dimension. By adding the two equations of motion
above and integrating the result two times we easily see that, for the chosen kink solution, the function $A(y)$
must be

\begin{equation}
A(y)=\frac{v^2}{72M^3}[4lncosh(ay)-tanh^2(ay)] \label{factor}
\end{equation}
where we have used $A(0)=0$ and $A'(0)=0$. Note that the exponential factor constructed with this function is
localized around the membrane and for large $y$ it looks like the Randall-Sundrum solution \cite{RS}. An
important characteristic of this solution is that for small perturbations of the metric it is possible to show
that there is a non-massive gravitational mode trapped to the membrane \cite{kehagias}. The solution found here
reproduces the Randall-Sundrum model in an specific limit. The spacetime now has no singularity because we get a
smooth warp factor (because of this, the model is more realistic). In fact this can be seen by calculating the
curvature invariants for this geometry. For example, we obtain

\begin{equation}
R=-[8A''+20(A')^{2}],
\end{equation}
where
\begin{equation}
A'(y)=-a\beta\tanh(ay)(2+\frac{1}{\cosh^{2}(ay)}),
\end{equation}
\begin{equation}
A''(y)=-\frac{3a^{2}\beta}{\cosh^{4}(ay)},
\end{equation}
and $\beta= \frac{v^2}{36M^3}$. Note that the Ricci scalar is finite and its behavior can be observed in Figure
(\ref{curv}).

The potential $V(\phi)$ must satisfy the equations of motion (\ref{mov1}) e (\ref{mov2}). Solutions for
$\phi(y)$ and $A(y)$ will be
\begin{equation}
V(\phi)=-6M^3[A''+4(A')^2]
\end{equation}
and its easy to see that this potential gets modified in order to support the solution proposed. Substituting
the expression for $A(y)$ and expressing the potential in terms of the scalar field we arrive at

\begin{equation}
V(\phi)=\frac{\lambda}{4}(\phi^{2}-v^{2})^{2}-\frac{\lambda}{108M^{3}}\phi^{2}(\phi^{2}-3v^{2})^{2}.
\end{equation}

In the limit where $M\rightarrow\infty$ the potential coincides with the standard double well potential.

\section{The Kalb-Ramond field}

In this section we study the behavior of the Kalb-Ramond field in the gravitational background described in the
last section. The main subject here is to try to detect a massless zero mode of the Kalb-Ramond field localized
on the membrane of the model described here. Note that the extra dimension has infinite size, an important
detail in order to achieve the final conclusion of this work. The action for the model is given by
\begin{equation}
S=\int d^{5}x \sqrt{-G}[2M^{3}R-\frac{1}{2}(\partial\phi)^{2}-V(\phi)-H_{MNL}H^{MNL}]
\end{equation}
where $H_{MNL}=\partial_{[M}B_{NL]}$ is the field strength for the Kalb-Ramond field. The equation of motion for
the field $B_{MN}$ is easily obtained.

First, we obtain the following equation:
\begin{equation}
\partial_{Q}(\sqrt{-G}H_{MNL}G^{MQ}G^{NR}G^{LS})=0.
\end{equation}
Now, we consider a gauge choice as $B_{\alpha5}=\partial_{\mu}B^{\mu\nu}=0$. Working with the part dependent on
the extra dimension, we can rewrite the equation of motion above as
\begin{equation}
e^{-2A}\partial_{\mu}H^{\mu\gamma\theta}-\partial_{y}H^{y\gamma\theta}=0.
\end{equation}
Here we can split the variables through the following ansatz:
\begin{equation}
B^{\mu\nu}(x^{\alpha},y)=b^{\mu\nu}(x^\alpha)U(y)=b^{\mu\nu}(0)e^{ip_\alpha x^\alpha}U(y),
\end{equation}
where $p^2=-m^2$. Therefore, we can rewrite $H^{MNL}$ as $h^{\mu\nu\lambda}U(y)$, and the equation of motion
becomes
\begin{equation}
\partial_{\mu} h^{\mu\nu\lambda}U(y)-e^{2A}\frac{d^2U(y)}{dy^2}b^{\nu\lambda}e^{ip_\alpha x^\alpha}=0.
\end{equation}
The function $U(y)$ carries all the information about the extra dimension and obeys the following equation:
\begin{equation}
\frac{d^2U(y)}{dy^2}=-m^2e^{-2A(y)}U(y).
\end{equation}

For the case where $m^{2}=0$ its solution is quite simple: $U(y)=cy+d$, where $c$ and $d$ are constants. Another
solution is $U(y)\equiv cte$.

Now it is important to analyze the effective action in $D=4$ for the Kalb-Ramond field. The way chosen for this
is based in the procedure of dimensional reduction. We focus our attention specifically on the massless zero
mode that appears from that reduction. For the metric described in the previous section we have the following:

\begin{equation}
S\sim\int \sqrt{-G}d^{5}x(H_{MNL}H^{MNL})=\int dy U(y)^{2}e^{-2A(y)}\int
d^{4}x(h_{\mu\nu\alpha}h^{\mu\nu\alpha}).
\end{equation}

\begin{figure}
\includegraphics[width=12cm]{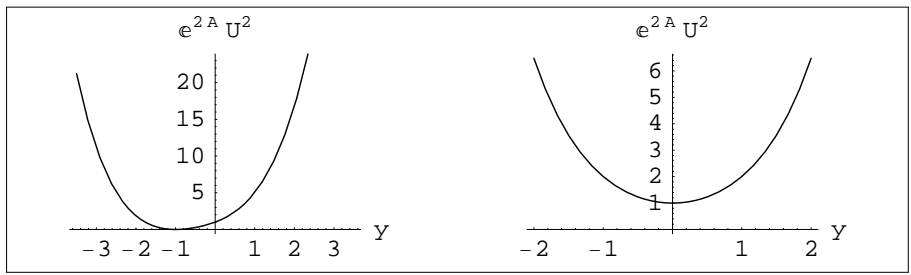}
\caption{
  \label{fig.1}  Plots of $U(y)^2 e^{2A(y)}$ with $A(y)$ given by
eq.(\ref{factor}) for $U(y)=y+1$ (left), $U(y)=1$ (right), $\beta=0.5$ and $a=1$. }
\end{figure}

Given the solution (\ref{factor}) for the warp factor $A(y)$ , we can see that for both type of solutions $U(y)$
obtained above, the integral in the $y$ variable in the effective action for the Kalb-Ramond zero mode is not
finite. This can be easily seen if we analyse the plot of the function of $y$ which is supposed to be
integrated. Indeed, for the solution $U(y)=cy+d$, and for the solution $U(y)\equiv cte$ as showed in Fig.
(\ref{fig.1}), the integral in $y$ do not converge.


Therefore, there is not localization of the Kalb-Ramond field on the membrane in the conditions described. This
result should not be a surprise: the gauge vector field suffers the same problem.

\section{Membrane with internal structures}

Due to the result of the last section, we may ask if there is some way to modify our model in order to change
the final conclusion. Models containing more fields may, in principle, give us different results because of its
natural and rich amount of information. Here we choose models that presents membranes with internal structures.
Membranes with internal structures can be described by models with two real scalar fields. Indeed, in Ref.
\cite{defects-inside}, Bazeia \textit{et. al.} introduced a model which supports Bloch walls, which in turn,
exist in ferromagnetic systems \cite{wal}, and already have internal structures. Following ideas where domain
walls are used in order to generate branes in scenarios where scalar fields couple with gravity, Bazeia and
Gomes proposed the so called Bloch Brane \cite{dionisio-adauto}. This brane model can be thought as an
alternative to the infinitely thin brane model, which, as pointed out in \cite{bonjour}, may be a very
artificial construction.

To the best of our knowledge, there are no previous study about localization of fields in that type of brane. It
is worthwhile to mention, however, a recent work by Gomes \cite{adalto}, which study \textbf{gravity}
localization in Bloch Branes.

\begin{figure}
\includegraphics[width=8cm]{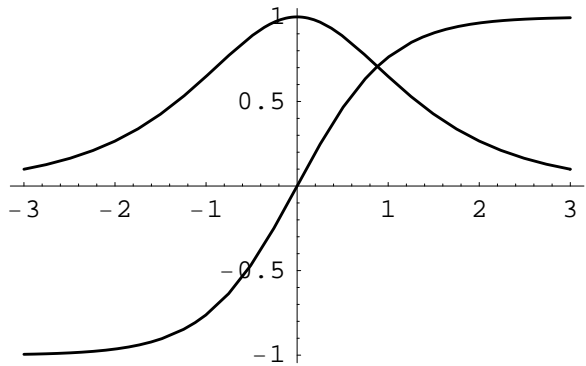}
\caption{
  \label{fig.7}   The kink-like object represents the membrane while
the lump-like object represents the internal structure.}
\end{figure}

We now study a model of two real scalar fields. The spacetime has the same characteristics as those discussed in
the second section: $ds^{2}=e^{2A(y)}\eta_{\mu\nu}dx^{\mu}dx^{\nu}+dy^{2}$. The question is the same: can this
model trap to a membrane a zero-mode of the Kalb-Ramond field? The difference here is that the membrane has
internal structures as explained above: we are discussing now a Bloch-type brane. We can see in Fig.
(\ref{fig.7}) the profile of the solution for a model of two real scalar fields.

We get a membrane which have an internal structure due to the scalar field denoted by $\chi$. This structure
gives rise to other defects (which are called domain ribbons) inside the membrane generated by the field $\phi$
\cite{defects-inside}. We must verify, therefore, if the existence of this internal structure may give us
additional information in order to localize the Kalb-Ramond gauge field.

The model is basically given by the following action:

\begin{equation}
S=\int d^4x dy \sqrt{-G}\bigg[\frac{1}{4}R+\frac{1}{2}\partial_a \phi
\partial^a \phi + \frac{1}{2}\partial_a \chi \partial^a
\chi-V(\phi,\chi)\bigg],
\end{equation}
where the metric is
\begin{equation}
ds^{2}=e^{2A(y)}\eta_{\mu\nu}dx^{\mu}dx^{\nu}+dy^{2}.
\end{equation}
The solutions we need to simulate the brane with internal structure are given by:
\begin{equation}
\phi(x)=tanh(2rx)
\end{equation}
\begin{equation}
\chi(x)=\pm\sqrt{\frac{1}{r}-2sech(2rx)}
\end{equation}


Following the procedure outlined in the previous sections we can easily find the warp factor as given by ($r$ is
a parameter of the model)
\begin{equation}
exp[2A(y)]=\cosh^{-4}(2ry)exp[\frac{2}{9r}(1-3r)\tanh^{2}(2ry)].
\end{equation}

For the effective action in $D=4$ of the Kalb-Ramond field we should take the inverse of this factor. Now, for
the function $U(y)$ constant, as discussed in the last section, we reobtain the result of non-localization of a
tensorial zero-mode. The Bloch brane does not change the setup for the Kalb-Ramond field.

\section{The dilaton}

In this section we continue to study models containing more than one scalar field. Due to the result of the last
section, it is natural to ask if there is some coupling between the Kalb-Ramond field and another field in order
to provide the localization of a tensorial zero mode. In analogy with the work of Kehagias and Tamvakis
\cite{kehagias}, where it is shown that the coupling between the dilaton and a vector gauge field produces
localization of the latter, we introduce here the coupling between the dilaton and the Kalb-Ramond field. The
coupling is motivated in low energy superstring theories \cite{mayr}.

Nevertheless, before analyzing the coupling, it is necessary to obtain a solution of the equations of motion for
the gravitational field in the background of the dilaton and the membrane. For such, we introduce the following
action \cite{kehagias}:
\begin{equation}
S=\int d^{5}x
\sqrt{-G}[2M^{3}R-\frac{1}{2}(\partial\phi)^{2}-\frac{1}{2}%
(\partial\pi)^{2}-V(\phi,\pi)].
\end{equation}
Note again that we are working with a model containing two real scalar fields. The field $\phi$ again plays the
role of to generate the membrane of the model while the field $\pi$ represents the dilaton. The potential
function now depends on both scalar fields. Furthermore, the Ricci scalar depends on a special form of the
dilaton field, as we shall see ahead. But, the key point here is that it is assumed a new ansatz for the
spacetime metric:
\begin{equation}
ds^{2}=e^{2A(y)}\eta_{\mu\nu}dx^{\mu}dx^{\nu}+e^{2B(y)}dy^{2}.
\end{equation}
The equations of motion are given by
\begin{equation}\label{eq1}
\frac{1}{2}(\phi^{\prime})^{2}+\frac{1}{2}(\pi^{\prime})^{2}-e^{2B(y)}V(%
\phi,\pi)=24M^{3}(A^{\prime})^{2},
\end{equation}
\begin{equation}\label{eq2}
\frac{1}{2}(\phi^{\prime})^{2}+\frac{1}{2}(\pi^{\prime})^{2}+e^{2B(y)}V(%
\phi,\pi)=-12M^{3}A^{\prime\prime}-24M^{3}(A^{\prime})^{2}+12M^{3}A^{\prime}B^{%
\prime},
\end{equation}
\begin{equation}\label{eq3}
\phi^{\prime\prime}+(4A^{\prime}-B^{\prime})\phi^{\prime}=\partial_{\phi}V,
\end{equation}
and
\begin{equation}\label{eq4}
\pi^{\prime\prime}+(4A^{\prime}-B^{\prime})\pi^{\prime}=\partial_{\pi}V.
\end{equation}

In order to solve that system, we use a so-called superpotential function $W(\phi)$, defined by
$\phi^{\prime}=\frac{\partial W}{\partial\phi}$, following the approach of Kehagias and Tamvakis
\cite{kehagias}. The particular solution regarded follows from choosing the potential $V(\phi,\pi)$ and
superpotential $W(\phi)$ as
\begin{equation}
V=e^{\frac{\pi}{\sqrt{12M^{3}}}}\{\frac{1}{2}(\frac{\partial
W}{\partial\phi}%
)^{2}-\frac{5}{32M^{2}}W(\phi)^{2}\},
\end{equation}
and
\begin{equation}
W(\phi)=va\phi(1-\frac{\phi^{2}}{3v^{2}}).
\end{equation}
Following the procedure it is easy to obtain first order differential equations whose solutions are solutions of
the equations of motion (\ref{eq1}-\ref{eq4}) above, namely
\begin{equation}\label{pi}
\pi=-\sqrt{3M^{3}}A,
\end{equation}
\begin{equation}\label{b}
B=\frac{A}{4}=-\frac{\pi}{4\sqrt{3M^{3}}},
\end{equation}
and
\begin{equation}
A^{\prime}=-\frac{W}{12M^{3}}.
\end{equation}
The solutions for these new set of equations are the following:
\begin{equation}
\phi(y)=v\tanh(ay),\label{dilat1}
\end{equation}
\begin{equation}
A(y)=-\beta \ln \cosh(ay)^{2}-\frac{\beta}{2}\tanh(ay)^{2},\label{dilat2}
\end{equation}
and
\begin{equation}
\pi(y)=\beta\sqrt{3M^{3}}(\ln \cosh(ay)^{2}+\frac{1}{2}%
\tanh(ay)^{2}).\label{dilat3}
\end{equation}
Following the argumentation in Ref. \cite{kehagias}, it is possible to see that, by the linearization of the
geometry described in this section, this model supports a massless zero mode of the gravitational field
localized on the membrane, even in the dilaton background.

\begin{figure}
\includegraphics[width=10cm]{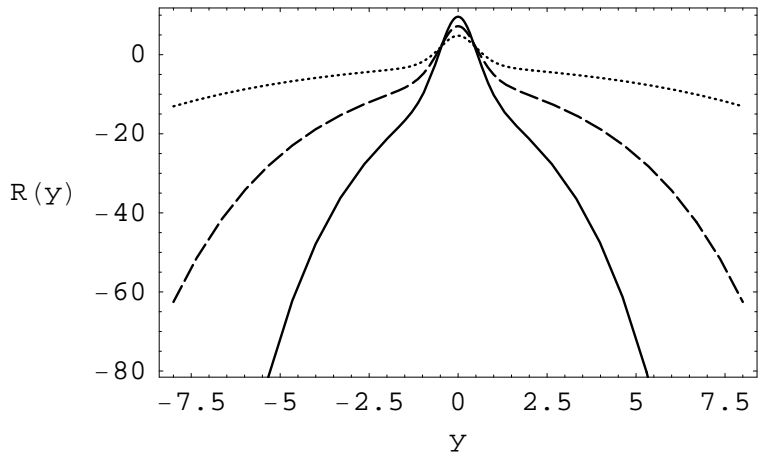}
\caption{
  \label{fig.3}   Plots of the solution of the curvature invariant
$R(y)$ with $\beta=0.07$ (dotted line), $0.1$ (dashed line), $0.14$ (solid line), $a=1$ and $3M^{3}=1$.}
\end{figure}

On the other hand the dilaton contribution makes the spacetime singular. However this kind of singularity is
very common in D-brane solutions in string theory (the dilaton solution is divergent). The Ricci scalar for this
new geometry is now given by

\begin{equation}
R=-[8A''+18(A')^{2}]e^{\frac{\pi}{2\sqrt{3M^{3}}}},
\end{equation}
where the dilaton has an important contribution. The new behavior of the Ricci scalar with the dilaton coupling
can be observed in Figure (\ref{fig.3}).

What is interesting here is that this singularity disappears if we lift the metric solution to $D=6$. In this
case the dilaton represents the radius of the sixth dimension \cite{kehagias}.

The next step is a review of the behavior of the Kalb-Ramond gauge field in this new background. Another
important point is the analysis of stability of the model. The mixing of scalar parts of all fields of this
model should be understood in order to see all aspects of localization. However it is subject for a next paper.
In this aspect some authors have already made some developments \cite{stability}.

It is worthwhile to point out that in scenarios where gravitational and scalar field fluctuations are considered, perturbations of the metric in transversal no trace sector splits from the scalar fields and then localized zero modes for the graviton are obtained. However, and we must emphasize this, despite the fact that in some models RS-type the Kalb-Ramond field could be included \textbf{with} the graviton, in our case the inclusion of the KR field is made through back reaction and therefore that field is \textbf{not} a constructive part of the geometry of the bulk.


\section{Reviewing the Kalb-Ramond field}

The main subject of this section is to verify if the coupling between the dilaton and the tensor gauge field is
able to produce localization of the latter. The dilaton coupling introduces the following modification in the
action for the tensor gauge field \cite{dilaton1,dilaton2}:
\begin{equation}
S\sim\int\sqrt{-G} d^{5}x(e^{-\lambda\pi}H_{MNL}H^{MNL}).
\end{equation}
Therefore, we must analyze the equations of motion of the tensor gauge field in the dilaton background. The new
equation of motion is:
\begin{equation}
\partial_{M}(\sqrt{-G}G^{MP}G^{NQ}G^{LR}e^{-\lambda\pi}H_{PQR})=0.
\end{equation}
With the gauge choice $B_{\alpha 5}=\partial\mu B^{\mu\nu}=0$ and with the separation of variables
$B^{\mu\nu}(x^{\alpha},y)=b^{\mu\nu}(x^{\alpha})U(y)=b^{%
\mu\nu}(0)e^{ip_{\alpha}x^{\alpha}}U(y)$ where $p^{2}=-m^{2}$, it is obtained the differential equation giving
us information about the extra dimension, namely
\begin{equation}
\frac{d^{2}U(y)}{dy^{2}}-(\lambda\pi^{\prime}(y)+B^{\prime}(y))\frac{dU(y)}{dy}=-m^{2}e^{2(B(y)-A(y))}U(y)\label{U1}.
\end{equation}
For the zero mode, $m=0$, a particular solution of the equation above is simply $U(y)\equiv cte$. This is enough
for the following discussion. The effective action for the zero mode in $D=5$ is
\begin{equation}
S\sim\int \sqrt{-G}d^{5}x(e^{-\lambda\pi}H_{MNL}H^{MNL})=\int dy U(y)^{2}e^{(-2A(y))+B(y)-\lambda\pi(y)}\int
d^{4}x(h_{\mu\nu\alpha}h^{\mu\nu\alpha}).
\end{equation}

For $U(y)$ as a constant and given the solutions for $A(y)$, $B(y)$ e $\pi(y)$, as in the last section, it is
possible to show clearly that the integral in the $y$ variable above is finite if $\lambda >
\frac{7}{4\sqrt{3M^3}}$, providing the possibility of localization of the zero mode associated to the
Kalb-Ramond field in the dilaton background.

\section{Kaluza-Klein massive modes}
In this section we study the massive spectrum of the Kalb-Ramond tensor field. This analysis allow us to detect
the presence of resonant mode solutions of the equation of motion dependent of the extra dimension. The dilaton
coupling was of great importance to get a localized zero-mode, as we have seen in Section (6). Then we will
follow our analysis of massive modes considering the same dynamics of dilaton coupling adopted in the previous
section. For this, initially we must transform the equation of motion for the extra dimension (\ref{U1}), in an
equation of the Schroedinger type.

Therefore we take the values of $\pi(y)$ and $B(y)$ to rewrite (\ref{U1}) in terms of $A(y)$ and its derivatives
on the following form,
\begin{equation}
\left\{\frac{d^2}{dy^2}-\alpha A'\frac{d}{dy}\right\}U(y)=-m^2 e^{-\frac{3}{2}A}U(y),
\end{equation}
were,
\begin{equation}
\alpha=\frac{1}{4}-\sqrt{3M^3}\lambda.
\end{equation}
In order to get an equation of the Schroedinger type from the equation above we must proceed with the following
mapping,
\begin{equation}\label{trans}
y\rightarrow z=f(y),\,\,\,\,\,\,  U=\Omega\overline{U}.
\end{equation}
The conditions to get an equation of the Schroedinger type must indicate the form of the function $\Omega$.
Therefore, we cannot have first derivative terms, and the right side of the required equation must contains the
constant $m^2$. We will have then,
\begin{equation}
\Omega=e^{\left(\frac{\alpha}{2}+\frac{3}{8}\right)A}, \,\,\,\,\,\, \frac{dz}{dy}=e^{-\frac{3}{4}A}.
\end{equation}
From the transformations (\ref{trans}), our Schroedinger-like equation can be written as,
\begin{equation}\label{schro}
\left\{-\frac{d^2}{dz^2}+\overline{V}(z)\right\}\overline{U}=m^2\overline{U},
\end{equation}
where the potential $\overline{V}(z)$ assumes the form,
\begin{equation}\label{pot_reson}
\overline{V}(z)=e^{\frac{3}{2}A}\left[\left(\frac{\alpha^2}{4}-
\frac{9}{64}\right)(A')^2-\left(\frac{\alpha}{2}+\frac{3}{8}\right)A''\right].
\end{equation}
We can write the potential in function of the derivatives respect to $z$,
\begin{equation}
\overline{V}(z)=\left[\beta^2(\dot{A})^2-\beta\ddot{A}\right]
\end{equation}
where,
\begin{equation}
\beta=\frac{\alpha}{2}+\frac{3}{8}
\end{equation}

The equation (\ref{schro}) will only present finite solution if we consider $\sqrt{3M^3}\lambda > 1$, which it
is compatible with the localization of zero-modes as described in the previous section.

\begin{figure}
\includegraphics[width=8cm]{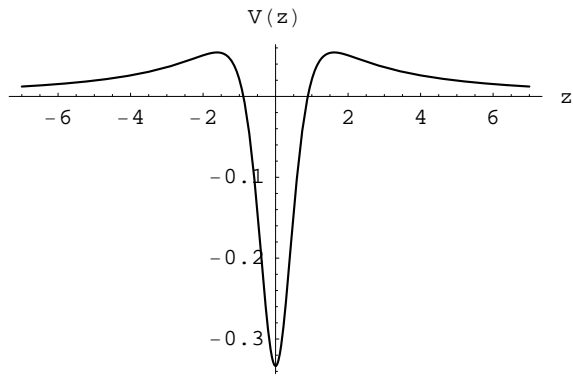}
\caption{
  \label{volkr}  Plots of potential $\overline{V}(z)$ were we put
$\frac{v^2}{72M^3}=1$}
\end{figure}

We plot the form of the potential in the Figure (\ref{volkr}), which suggests the possibility of resonant modes.
As we can note in Eq. (\ref{pot_reson}), the potential is such that $\overline{V}(z)\rightarrow 0$ when
$z\rightarrow\infty$. This excludes the possibility of gaps in the continuous spectrum.

It is interesting to point out that the Schroedinger-type equation (\ref{schro}) can be written in the
supersymmetric quantum mechanics scenario as follows,
\begin{equation}\label{susy_qm}
Q^{\dag} \, Q \,
\overline{U}(z)=\left\{\frac{d}{dz}-\beta\dot{A}\right\}\left\{\frac{d}{dz}+\beta\dot{A}\right\}\overline{U}(z)=-m^2\overline{U}(z)
\end{equation}
From the form of the Eq. (\ref{susy_qm}), we exclude the possibility of normalized negative energy modes
existence. On the other hand, we exclude also the possibility of the presence of tachyonic modes, which is a
necessary condition to keep the stability of gravitational background.

\begin{figure}
\includegraphics[width=8cm]{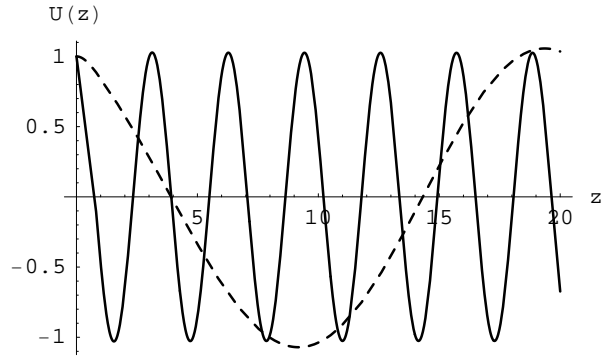}
\caption{
  \label{wave}  Plots of wave function for $m^2=4$ (solid line), and
$m^2=0.1$ (dashed line), where we put $\frac{v^2}{72M^3}=1$.}
\end{figure}

We cannot find analytical solution of the massive modes wave function in Schroedinger equation. However we will
be able to analyze the solution for $\overline{U}$ by numerically solving the equation (\ref{schro}). We plot in
Figure (\ref{wave}) the wave function so obtained for two values of $m^2$. As we can observe, the wave function
oscillates quickly for a moderate value of $m^2$ and reduces its period for small values of $m^2$. As mentioned
in Ref. \cite{dionisio-adauto}, this behavior of the wave function suggests a free motion in the bulk, but no
imprisonment in the membrane.

\begin{figure}
\includegraphics[width=8cm]{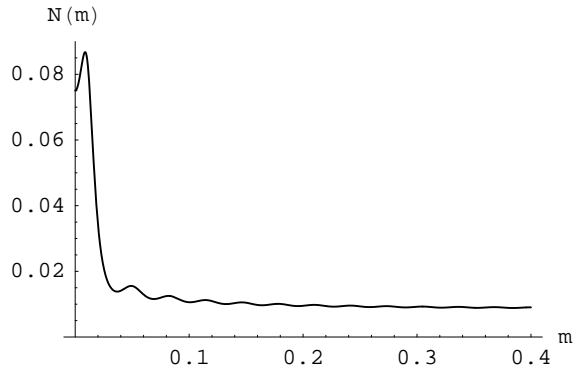}
\caption{ \label{reso} Plots of $N(m)$ with $\sqrt{3M^3}\lambda = 2$}
\end{figure}

In a close resemblance to the analysis of Refs. \cite{csaba,gremm}, for some values of mass, our plane wave
solutions can assume very high amplitudes inside the brane and this may be understood in terms of resonance
structures. In figure (\ref{wave}) we analyze only two modes in which we observe only variations in the period
of the oscillations without great values for $\overline{U}(z)$ at $z=0$. Therefore these two modes do not
present resonances. However, as mentioned in Ref. \cite{csaba2}, in order to detect the presence of resonances
we must know the value of our solutions inside the membrane as function of the mass. In this way we will be
searching a great number of values of mass capable to produce great amplitudes in $z=0$.

The equation for massive modes written in the form of Eq.(\ref{susy_qm}) allows us to interpret
$|\chi\overline{U}(z)^2|$ as a probability to find the mode in the position $z$, where $\chi$ is a normalization
constant. Therefore, we must calculate $|\chi\overline{U}(0)^2|$ as a function of the mass in order to determine
the intensity of the modes on the brane.

First, we solve numerically the Eq. (\ref{schro}) for a sequence of mass values in the interval $0<m<1$ with a
steep of $10^{-3}$. We have chosen this interval because the potential cause just a little perturbation for the
modes for which $m^2>>V_{max}$ \cite{gremm}). Hence, in the case of existing a resonant structure, we hope to
find it for $0<m<0.24$.

In order to normalize our plane wave function, we restrict the solutions of each mode to the region $-100<z<100$
and we extract a normalization constant $\chi$ for each correspondent value of mass. Then, we interpolate our
data bank and we construct the function $N_m=|\chi\overline{U}_0(m)^2|$ which give us the probability to find
the modes in $z=0$ as function of $m$. We plot $N(m)$ in the figure (\ref{reso}). As we can see, there is a
resonance peak near $m=0$, or more precisely, at $m=0.009$. The analysis of the graphic (\ref{reso}) shows that
for $m=0$ we have a big value for $N(m)$, assuring that we have a zero mode localization.

On the other hand, the coupling of the modes with matter in the brane is related to the normalized wave function
amplitude in $z=0$ \cite{gremm}. In our case, we impute that relation to the function $N(m)$. As we expected,
non-massive modes show a high value for $N(m)$, considering that they are coupled to the matter in the brane.
The resonance in $m=0.009$ shows us that only very light modes couple to the matter in a magnitude of the order
of that of the zero modes. Furthermore, as we can see in Fig. (\ref{reso}), for heavy modes, $N(m)$ tends to a
constant value. Therefore, we can conclude that these modes have a weak coupling, if we compare they with the
coupling of the very light ones.

An interesting additional point consists in testing the consistency of the function $N(m)$ taking in to account
the presence of the dilaton. In section III, when we consider the brane without internal structures, the zero
mode of the KR field was not localized. Hence, if we consider that scenario, the function $F(m)$ must be
changed. In the examples discussed above, we have used $\sqrt{3M^3}\lambda = 2$. Therefore, in order to make the
dilaton coupling disappears, we must to use $\lambda=0$ and recalculate the function $N(m)$. As we can note the
zero mode coupling is highly suppressed compared to the massive modes and the resonance disappears. That result
agreed with our results of the section III, namely, if we do not consider the dilaton coupling we obtain  non
localized modes again.



\begin{figure}
\includegraphics[width=8cm]{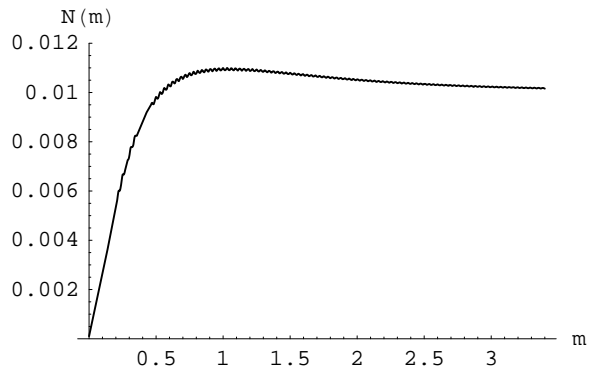}
\caption{ \label{reso2} Plots of $N(m)$ with $\lambda =0$}
\end{figure}

\section{Conclusions}
In this work we have made a study of localization of zero and massive modes of a Kalb-Ramond gauge field on
several types of branes. In a first case, we have just seen the contribution of the spacetime geometry in the
behavior of a possible localized zero-mode. We have chosen an AdS-like geometry generated by a model of real
scalar fields in five dimensions. We have shown that the solutions for this case does not guarantee the presence
of zero modes trapped to the membrane. The reason for such a behavior is due to the infinite size of the extra
dimension which makes the effective action for the Kalb-Ramond field in $D=4$ strongly divergent. Stepping
further, we have added the dilaton field through a coupling to the Kalb-Ramond field in order to evaluate if
this new ingredient produces normalized zero modes. The answer we have found is positive within certain
conditions over the coupling constant $\lambda$, i.e, the dilaton contributes to localize such a zero mode, as
in the case for the vector gauge field already discussed in Literature.

After this we address the issue of zero mode confinement in thick branes, in particular those called Bloch
branes, which exhibit an internal structure, based in two real scalar fields and we have found again a negative
answer for the localization of the Kalb-Ramond zero mode. Nevertheless, in the presence of the dilaton
background we have zero mode localization. In a certain way, the results for the Kalb-Ramond gauge field without
the dilaton background are corroborated by the analogous study in the Randall-Sundrum scenario where the extra
dimension is compacted in an orbifold \cite{sengupta}. In that case there is a zero mode but strongly suppressed
by the size of the extra dimension. Effectively, in the Randall-Sundrum scenario there is no dynamical
Kalb-Ramond gauge field in $D=4$. Finally, we study the Kaluza-Klein massive modes of Kalb-Ramond field. Using
an approach of supersymmetric quantum mechanics in a Schrodinger-like equation, we exclude the possibility of
gaps in the continuous spectrum.

By numerically solving the Schrodinger like equation of motion, initially for two modes, we get solutions of
plane-waves, an indication of non localization. Knowing the variation of our solutions in the brane as function
of mass, we detect the presence of a resonant modes. Therefore, from an analysis of the resonance structure we
conclude that heavy massive modes have a weak coupling, if we compare they with the coupling of the very light
ones. In addition, if we consider a situation where the coupling of the dilaton vanishes, the resonance
disappears and the zero mode is highly suppressed compared to the massive modes.

The authors would like to thank Funda\c{c}\~{a}o Cearense de apoio ao Desenvolvimento Cient\'{\i}fico e
Tecnol\'{o}gico (FUNCAP) and Conselho Nacional de Desenvolvimento Cient\'{\i}fico e Tecnol\'{o}gico (CNPq) for
financial support. Also, we are indebted to anonymous referees whose comments have helped us a lot to improve
the manuscript.

\end{document}